\documentclass[11pt,twoside]{article}

\usepackage{asp2004}
\usepackage{epsf}
\usepackage{psfig}
\usepackage{lscape}

\def\HI   {H~{\small I}}
\def\HII  {H~{\small II}}
\def\NII  {[N~{\small II}]}
\def\NeII {[Ne~{\small II}]}
\def\NeIII{[Ne~{\small III}]}
\def\NeV  {[Ne~{\small V}]}
\def\OI   {[O~{\small I}]}
\def\OII  {[O~{\small II}]}
\def\OIII {[O~{\small III}]}
\def\OIV  {[O~{\small IV}]}
\def\SII  {[S~{\small II}]}
\def\SIII {[S~{\small III}]}

\def\SiII {[Si~{\small II}]}
\def\m{~$\mu$m}
\def\ISO{{\it ISO}}

\def\Spitzer{{\it Spitzer}}
\def\be{\begin{equation}}
\def\ee{\end{equation}}

\begin{document}
\title{Mid-Infrared Spectral Diagnostics of Nearby Galaxies}   %%% Fill in title
\author{Daniel A. Dale \& the SINGS Team}   %%% Fill in author names
\affil{University of Wyoming}    %%% Fill in author affiliations

\begin{abstract} 
The {\it Spitzer Space Telescope} is pushing into new frontiers in high redshift astronomy.  Closer to home, {\it Spitzer} is making an equally large impact on our understanding of galaxy formation and evolution.  In this contribution we present mid-infrared diagnostics based largely on data from the Spitzer Infrared Nearby Galaxies Survey (SINGS).  Our main result is that these mid-infrared diagnostics effectively constrain a target's dominant power source.  The combination of a high ionization line index and PAH strength serves as an efficient discriminant between AGN and star-forming nuclei, confirming progress made with {\it ISO} spectroscopy on starbursting and ultraluminous infrared galaxies.  The sensitivity of {\it Spitzer} allows us to probe fainter nuclei and star-forming regions within galaxy disks.  We find that both star-forming nuclei and extranuclear regions stand apart from nuclei that are powered by Seyfert or LINER activity.  In fact, we identify areas within three diagnostic diagrams containing $>$90\% Seyfert/LINER nuclei or $>$90\% \HII\ regions/\HII\ nuclei.  We also find that, compared to starbursting nuclei, extranuclear regions typically separate even further from AGN, especially for low-metallicity extranuclear environments.  In addition, instead of the traditional mid-infrared approach to differentiating between AGN and star-forming sources that utilizes relatively weak high-ionization lines, we show that strong low-ionization cooling lines of X-ray dominated regions like \SiII~34.82\m\ can alternatively be used as excellent discrimants.  
\end{abstract}

\section {Introduction}
\label{sec:intro}

The goal of this study is to explore whether mid-infrared diagnostics developed for luminous/ultraluminous infrared galaxies and bright Galactic \HII\ regions can be improved upon and extended to the nuclear and extra-nuclear regions within normal and infrared-faint galaxies.  A traditional method for characterizing a galaxy's nuclear power source uses ratios of optical emission lines such as \OII~3727\AA, H$\beta$~4861\AA, \OIII~5007\AA, \OI~6300\AA, H$\alpha$~6563\AA, \NII~6584\AA\, and \SII~6716,6731\AA\ (e.g., Baldwin, Phillips, \& Terlevich 1981; Veilleux \& Osterbrock 1987; Ho, Filippenko, \& Sargent 1997; Kewley et al. 2001; Kauffmann et al. 2003).  A plot of \OIII/H$\beta$ versus \NII/H$\alpha$, for example, will typically separate Seyferts, LINERs, and starburst nuclei.  Since nuclei are often heavily enshrouded by dust, especially in luminous and ultraluminous infrared galaxies (LIRGs and ULIRGs), an important limitation to galaxy optical diagnostics is the effect of extinction.  In anticipation of the data stream from space-based infrared platforms, early theoretical work with photoionization models showed that infrared ionic fine structure line ratios could profitably enable astronomers to approach galaxy classification from a new perspective (e.g., Voit 1992; Spinoglio \& Malkan 1992).  The advent of sensitive infrared line data from the {\it Infrared Space Observatory} was an important first step to peering more deeply into buried nuclear sources (Genzel et al. 1998; Laurent et al. 2000; Sturm et al. 2002; Peeters, Spoon, \& Tielens 2004).  Genzel and collaborators were the first to show that ionization-sensitive indices based on mid-infrared line ratios correlate with the strength of polycyclic aromatic hydrocarbon (PAH) emission features.  AGN in particular show weak PAH and large ratios of high-to-low ionization line emission.  High ionization lines are not the only route, though, to determining whether a galaxy harbors a strong AGN.  Similar to how the \OI~6300\AA\ line is an AGN diagnostic, the \OI~63\m\ far-infrared line can also help to decipher a galaxy's power source (Dale et al. 2004a).  Mid-infrared lines observed by \ISO\ were also used to probe the physical characteristics and evolution of purely starbursting nuclei (Thornley et al. 2000; Verma et al. 2003) and Galactic \HII\ regions (Vermeij \& van der Hulst 2002; Giveon et al. 2002).  An important result stemming from these efforts is that stellar aging effects appear to result in \HII\ regions generally having higher excitations than starbursting nuclei.

The unprecedented sensitivity and angular resolution afforded by the {\it Spitzer Space Telescope} allow even more detailed studies of galaxy nuclei (e.g., Armus et al. 2004; Smith et al. 2004).  SINGS takes full advantage of {\it Spitzer}'s capabilities by executing a comprehensive, multi-wavelength survey of 75 nearby galaxies spanning a wide range of morphologies, metallicities, luminosities, and star formation activity levels (Kennicutt et al. 2003).  The sensitivity of {\it Spitzer} coupled with the proximity of the SINGS sample allows dwarf galaxy systems fainter than $L_{\rm FIR} \sim 10^7~L_\odot$ to be spectroscopically probed in the infrared for the first time.  In addition, prior to {\it Spitzer} the only individual extragalactic \HII\ regions that were detectable with infrared spectroscopy resided in the Local Group (e.g., Giveon et al. 2002; Vermeij et al. 2002).  In contrast, SINGS provides infrared spectroscopic data for nearly 100 extragalactic \HII\ regions, residing in systems as near as Local Group members to galaxies as far as $\sim$25~Mpc.  The SINGS dataset thus samples a wider range of environments than previously observed with infrared spectroscopy.  This diversity in the SINGS sample provides a huge range for exploring physical parameters with the mid-infrared diagnostics presented below.

\section {The Sample and Data}

The sample of nuclear targets analyzed in this study derive from the SINGS Third Data Release.  These 50 nuclei come from a wide range of environments and galaxies: low-metallicity dwarfs; quiescent ellipticals; dusty grand design spirals; Seyferts, LINERs, and starbursting nuclei of normal galaxies; and systems within the Local and M~81 groups (see Kennicutt et al. 2003).  The 26 extranuclear sources studied in this work also come from the SINGS Third Data Release.  These targets stem from the original set of 39 optically-selected sources listed by Kennicutt et al. (2003).  The optically-selected OB/\HII\ regions cover a large range of metallicity, extinction-corrected ionizing luminosity, extinction, radiation field intensity, ionizing stellar temperature, and local H$_2$/\HI\ ratio as inferred from CO.  Further details on these data and the archival data used in this study are fully described in Dale et al. (2006).

\subsection {Mid-Infrared Spectral Diagnostics}
\subsubsection {Emission Line Ratios and PAH Strength}
\label{sec:lines_pah}

\begin{figure}
 \plotone{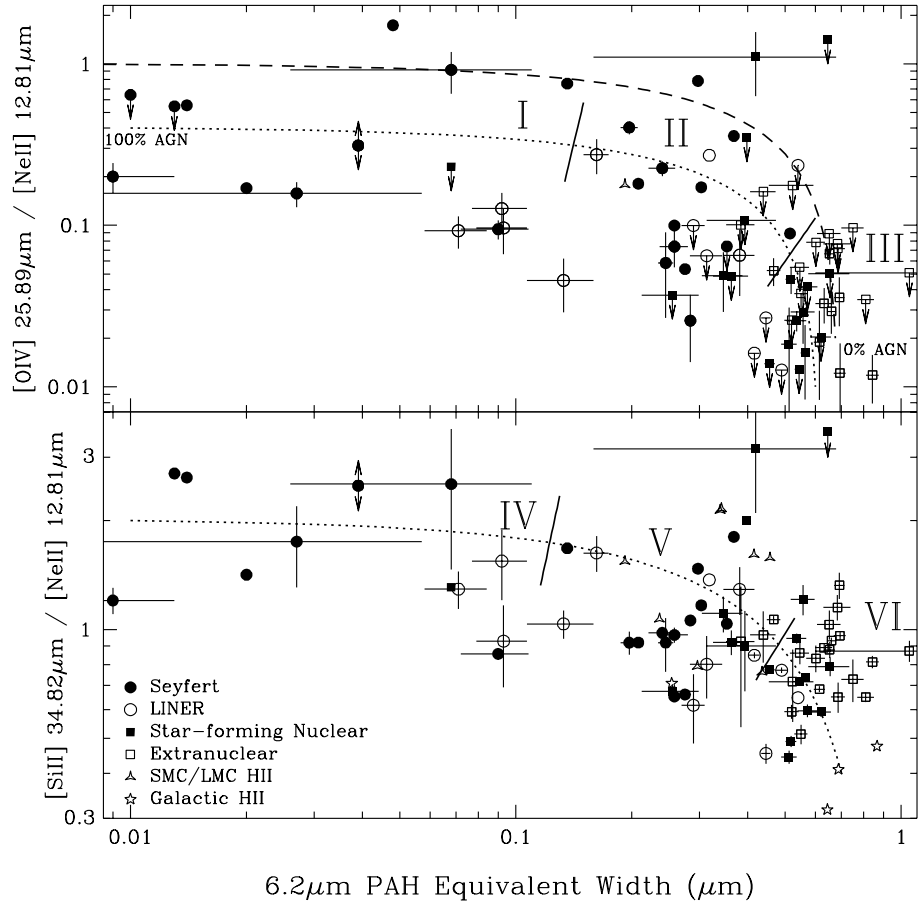}
 \caption{The ratios of mid-infrared forbidden lines 
 %of the \OIV\ and \NeII\ lines and the \SiII\ and \NeII\ lines 
 as a function of the 6.2\m\ PAH feature equivalent width.  SINGS data are displayed with 1$\sigma$ error bars based on the statistical uncertainties; archival data are displayed without error bars.  The dotted lines are linear mixing models of a ``pure'' AGN and a ``pure'' star-forming source (see text).  The dashed line in the top panel is a similar mixing model first presented by Genzel et al. (1998).  The solid lines delineate regions distinguished by Seyferts, LINERs, star formation, etc. (see \S~\ref{sec:lines_pah}).}
 \label{fig:lines_pah}
\end{figure}

Studies show that PAH features are quite prominent throughout much of the interstellar medium except for regions characterized by exceptionally hard radiation fields such as those that arise in AGN and the cores of \HII\ regions (Cesarsky et al. 1996; Sturm et al. 2000).  Furthermore, the different energies required to stimulate \OIV 25.89\m\ and \NeII 12.81\m\ means their flux ratio depends on the type of source dominating the energetics of the interstellar medium.  A mid-infrared diagnostic diagram first put forth by Genzel et al. (1998), and later explored by Peeters, Spoon, \& Tielens (2004), plots the emission line ratio \OIV 25.89\m/\NeII 12.81\m\ versus the strength of a mid-infrared PAH feature.  In such plots AGN sources show enhanced \OIV 25.89\m\ emission and comparatively weak PAH feature strength.  However, those two studies focussed on AGN-dominated and starburst nuclei.  The upper panel of Figure~\ref{fig:lines_pah} uses the 6.2\m\ PAH feature and the emission line ratio \OIV 25.89\m/\NeII 12.81\m\ in a similar mid-infrared diagnostic diagram, but one that incorporates ``normal'' (starbursting/star-forming) nuclei and \HII\ regions---the sensitivity of \Spitzer\ allows us to probe to far fainter levels than heretofore possible.  As expected, normal nuclei and \HII\ regions extend the previously-observed trend: lower-luminosity star-forming nuclei and \HII\ regions exhibit comparatively large 6.2\m\ equivalent widths and relatively low ratios of \OIV 25.89\m/\NeII 12.81\m, indicating strong contributions from \NeII 12.81\m\ cooling of \HII\ regions and their PAH-rich photodissociation region surroundings, and negligible emission from AGN.  

High ionization lines like \OIV 25.89\m\ are somewhat difficult to detect in many SINGS sources.  An alternative diagnostic diagram is provided in the lower panel of Figure~\ref{fig:lines_pah} involving the more easily detectable \SiII\ 34.82\m\ line, 
which has an ionization potential of 8.15~eV.  As pointed out by Maloney, Hollenbach, \& Tielens (1996), the \SiII34.82\m\ line is a strong coolant of X-ray irradiated gas.  In X-ray dominated regions the X-ray emission dominates that from the comparatively small \HII-like regions surrounding the hard-spectrum source.  Moreover, X-ray dominated regions can be quite large since hard X-ray photons penetrate large column densities, and the conversion of X-ray energy to infrared continuum and line emission can be very efficient.  Maloney, Hollenbach, \& Tielens (1996) predict \SiII34.82\m\ to be prominent cooling line within X-ray dominated regions.  We take advantage of this concept to present a new technique for distinguishing between AGN sources and star-forming regions.  This technique relies on an easily-detectable, prominent cooling line of a low-ionization species associated with X-ray dominated regions, the dense interstellar material illuminated by power-law radiation fields.

The dotted lines in Figure~\ref{fig:lines_pah} represent a variable mix of an AGN nucleus and a ``pure'' star-forming region.  The 100\% AGN anchor for these mixing models have values of EW(PAH~6.2\m)$=0.01$\m, \OIV 25.89\m/\NeII 12.81\m=0.3, and \SiII 34.82\m/\NeII 12.81\m=2, whereas the 100\% star formation anchor has EW(PAH~6.2\m)$\sim$0.7\m, \OIV 25.89\m/\NeII 12.81\m=0.01, and \SiII 34.82\m/\NeII 12.81\m=0.4.  The dashed line in the upper panel of Figure~\ref{fig:lines_pah} shows the approximate mixing model of Genzel et al. (1998; their Figure~5), obtained after empirically deriving a relation between their 7.7\m\ PAH `strength' (line-to-continuum ratio) and the 6.2\m\ PAH equivalent width.  Many of the high ionization line data presented by Genzel et al. (1998) were upper limits, so it is unsurprising that their original curve lies above our curve.  Short solid lines roughly perpendicular to the dotted AGN/star-forming curves delineate three regions in both panels of Figure~\ref{fig:lines_pah}.  The boundaries are:
\be
{\rm Region [~I-~II]}: \log ( [{\rm O~IV}]25.89\micron / [{\rm Ne~II}]12.81\micron ) = 10~\log ({\rm EW}[6.2\micron {\rm PAH}]) + 8.0
\ee
\be
{\rm Region [II-III]}: \log ( [{\rm O~IV}]25.89\micron / [{\rm Ne~II}]12.81\micron ) = 3.4 \log ({\rm EW}[6.2\micron {\rm PAH}]) - 0.2
\ee
\be
{\rm Region [IV-V]}: \log ( [{\rm Si~II}]34.82\micron / [{\rm Ne~II}]12.81\micron ) = 5.0 \log ({\rm EW}[6.2\micron {\rm PAH}]) + 4.8
\ee
\be
{\rm Region [V-VI]}: \log ( [{\rm Si~II}]34.82\micron / [{\rm Ne~II}]12.81\micron ) = 1.7 \log ({\rm EW}[6.2\micron {\rm PAH}]) + 0.5
\ee
The population statistics for these regions show that Regions~[I+IV] and [III+VI] are respectively representative (at the $\sim$90\% level) of Seyferts and star-forming systems.  Regions [II+V], on the other hand, contain a mix of classifications and thus represent transition regions---either the source classifications in this region are ambiguous or the region simply contains a more heterogeneous mixture of pure types.

\subsubsection {A Neon, Sulfur, and Silicon Diagnostic}
\label{sec:NeSSi}

\begin{figure}
 \plotone{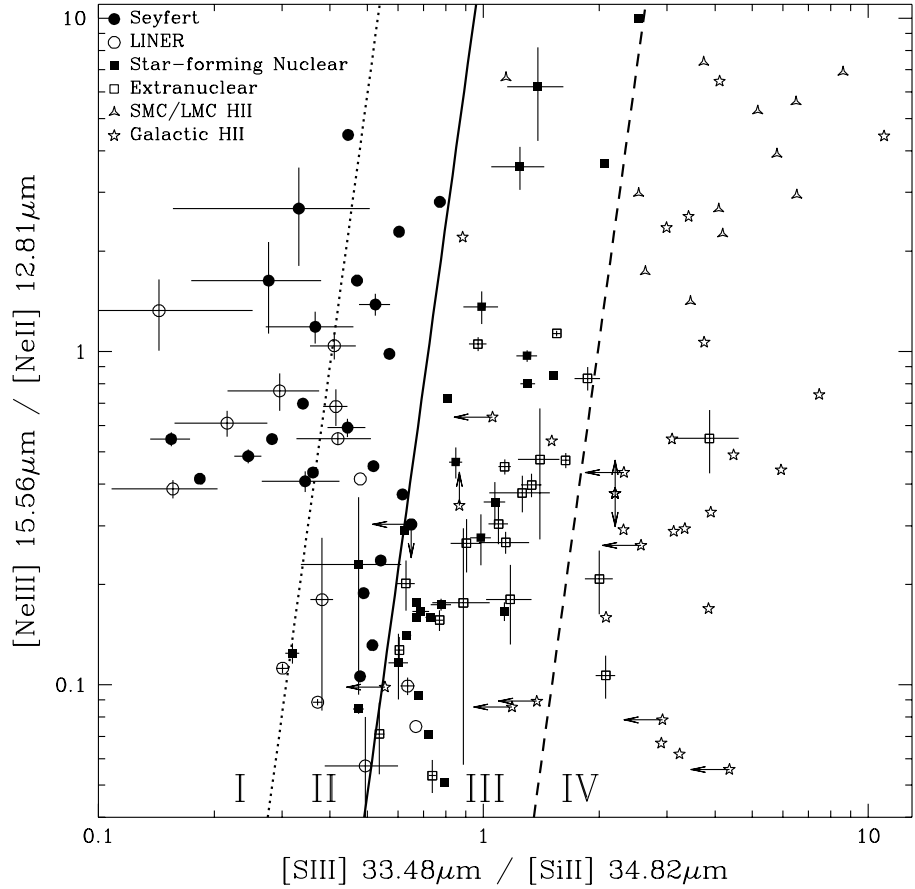}
 \caption{A neon, sulfur, and silicon diagnostic diagram involving ratios of lines at different ionizations is displayed.  The data are displayed as described in Figure~\ref{fig:lines_pah}.  The lines delineate regions distinguished by Seyferts, LINERs, star formation, etc. (see \S~\ref{sec:NeSSi}).}
 \label{fig:NeSSi}
\end{figure}

If the neon excitation is plotted as a function of \SIII 33.48\m/\SiII 34.82\m\ (Figure~\ref{fig:NeSSi}), a more obvious separation of the star-forming and AGN-powered data points is observed.  Not only do the low-metallicity Magellanic Cloud regions exhibit a higher neon excitation, nearly all of the ``pure star-forming'' nuclei and extra-nuclear regions show relatively elevated ratios in \SIII 33.48\m/\SiII 34.82\m.  Note in addition that many of the filled squares representing starbursting/star-forming nuclei are located between the \HII\ regions and the AGN.  %Table~\ref{tab:NeSSi} quantifies the source type fractions within each of the four regions delineated by the lines drawn in Figure~\ref{fig:NeSSi}.  
The boundaries are defined by curves with the same slope but differing offsets:
\be
\log ( [{\rm Ne~III}]15.56\micron / [{\rm Ne~II}]12.81\micron ) = 8.4 \log ( [{\rm S~III}]33.48\micron / [{\rm Si~II}]34.82\micron ) + \gamma,
\ee
where $\gamma=$[$+$3.3,$+$1.2,$-$2.5] for the lines demarcating Regions~[I$-$II,II$-$III,III$-$IV].

These results can be partially understood in the context of the cooling line physics introduced above.  The \SiII 34.82\m\ line is a significant coolant of X-ray ionized regions or photodissociation regions (Hollenbach \& Tielens 1999), whereas the \SIII 33.48\m\ line is a strong marker of \HII\ regions.  In other words, extra-nuclear regions and star-forming nuclei will show strong signatures of the Str\"omgren sphere coolant \SIII 33.48\m, while AGN and their associated X-ray dominated regions or dense photodissociation regions will exhibit relatively strong \SiII 34.82\m\ emission in analogy to the increased strength of \OI~6300\AA\ emission in AGN (e.g., Veilleux \& Osterbrock 1987).  In addition, the fraction of photodissociation regions falling within each beam will play a role in the line ratios.  The data for Magellanic Cloud and Galactic \HII\ regions stem from smaller physical apertures and thus are likely to have fractionally higher contributions from Str\"omgren spheres than photodissociation regions.

Metallicity may be a factor as well.  Since the central regions of galaxies typically are more abundant in heavy metals (Pagel \& Edmunds 1981; McCall 1982; Vila-Costas \& Edmunds 1992; Pilyugin \& Ferrini 1998; Henry \& Worthey 1999), and as explained above a lower metallicity can lead to harder radiation fields and thus enhanced high-ionization-to-low-ionization line ratios, it is possible that this AGN$\rightarrow$\HII\ nucleus$\rightarrow$\HII\ region sequencing along the \SIII 33.48\m/\SiII 34.82\m\ axis is affected by metallicity.  However, the lower metallicity Magellanic Cloud data are not substantially to the right of the Galactic \HII\ region data, so the effect is not solely due to metallicity.  Alternatively, perhaps some of the star-forming nuclei have contributions from undetected weak AGN and thus are not ``pure'' star-forming nuclei, resulting in a location for star-forming nuclei on this diagram between AGN and \HII\ regions.

\section {Summary}

We have presented mid-infrared diagnostic diagrams for a large portion of the SINGS sample supplemented by archival \ISO\ and \Spitzer\ data.  A portion of our work solidifies and extends previous \ISO-based mid-infrared work to lower luminosity normal galaxy nuclei and \HII\ regions using the {\it Spitzer Space Telescope}.  We also present new diagnostics that effectively constrain a galaxy's dominant power source.  The power of the diagnostic diagrams of Genzel et al. (1998; see also Peeters, Spoon, \& Tielens 2004) for distinguishing between AGN and star-forming sources in dusty ULIRGs is that mid-infrared lines and PAH features are much less affected by extinction than their optical counterparts in a traditional diagnostic diagram.  Unlike diagrams put forth by Genzel et al., which rely on detecting relatively weak high-ionization lines like \OIV~25.89\m\ and \NeV~14.32\m, we provide a new diagnostic that utilizes a strong {\it low}-ionization line.  The advantage of using a line ratio like \SiII/\NeII\ is that singly-ionized silicon and neon respectively have ionization potentials of only 8.15 and 12.8~eV, so they can both be observed over a large range of physical conditions.  This is similar in concept to previous efforts that have taken advantage of \OI\ lines (e.g., at 6300~\AA\ or 63\m) that are coolants of the X-ray dominated regions (or dense photodissociation regions) surrounding AGN.  In plots of both \OIV/\NeII\ vs PAH equivalent width and \SiII/\NeII\ vs. PAH equivalent width we identify a region where $>$90\% of the sources are Seyfert or LINER.  Likewise, another region in both plots is $>$90\% \HII\ regions or star-forming nuclei.

Another useful mid-infrared diagnostic is \NeIII~15.56\m/\NeII~12.81\m\ vs. \SIII~33.48\m/\SiII~34.82\m.  This plot tracks the excitation power of the radiation field on one axis, while the other axis is a relative measure of the cooling of \HII\ regions and X-ray dominated regions (or dense photodissociation regions).  Similar to what is found for the \SiII/\NeII\ vs. PAH diagnostic alluded to above, both starbursting nuclei and extranuclear regions stand apart from nuclei that are powered by accretion-powered disks.  Moreover, compared to starbursting nuclei, extranuclear regions typically separate even further from Seyfert nuclei, especially for low-metallicity environments.  Presumably this extranuclear$\longleftrightarrow$nuclear separation occurs since extranuclear regions are cleaner representatives of \HII\ regions than starburst nuclei, because their stellar populations and interstellar medium structure are less complex.  Extranuclear regions more likely contain younger stellar populations since they trace a single burst, as opposed to the average of multiple star formation episodes for nuclei (e.g., Dale et al. 2004b).  Finally, we note that it is difficult to clearly distinguish between pure Seyfert and pure LINER sources using these diagnostics.

\acknowledgements 
Support for this work, part of the {\it Spitzer Space Telescope} Legacy Science Program, was provided by NASA through Contract Number 1224769 issued by the Jet Propulsion Laboratory, California Institute of Technology under NASA contract 1407.

%FIGURES
%%%%%%%%%%%%%%%%%%%%%%%%%%%%%%%%%%%%%%

%%%%%%%%%%%%%%%%%%%%%%%%%%%%%%%%%%%
\end{document}